\documentclass[11pt,letterpaper]{article}
\usepackage{soul}
\usepackage{jheppub}
\usepackage[utf8]{inputenc}
\usepackage[T1]{fontenc}
\usepackage{amsfonts}
\usepackage{tikz}
\usepackage{amsmath}
\usepackage{graphicx,subfigure}
\usepackage{tensor} 
\usepackage{mathtools} 
\usepackage{amssymb}
\usepackage{enumerate}
\usepackage{euscript}
\usepackage{mathrsfs}
\usepackage{hyperref}
\usepackage{indentfirst}
\usepackage{color}
\usepackage{float}
\usepackage{braket}
\title{\boldmath Spontaneous Symmetry Breaking and the Emergent Einstein-Standard Model: From $\text{Weyl} \times SU(2)_L\times U(1)_Y$ Gauge Theory to Geometric Mass Generation}

\author[]{Hao-Qian Peng$^{ab}$,}
\author[]{Yun-Tao Gu$^{ab}$,}
\author[]{Yu-Xiao Liu$^{ab}$\footnote{Corresponding author}}
\affiliation[a]{
  Lanzhou Center for Theoretical Physics, Key Laboratory of Theoretical Physics of Gansu Province, Key Laboratory of Quantum Theory and Applications of MoE, Gansu Provincial Research Center for Basic Disciplines of Quantum Physics, Lanzhou University, Lanzhou 730000, China
}
\affiliation[b]{Institute of Theoretical Physics \& Research Center of Gravitation,
School of Physical Science and Technology, Lanzhou University, Lanzhou 730000, China}

\emailAdd{penghq2024@lzu.edu.cn,guyt2024@lzu.edu.cn,liuyx@lzu.edu.cn}

\abstract{We construct a 
$\text{Weyl} \times SU(2)_L\times U(1)_Y$ invarient theory by extending four-dimensional Weyl quadratic gravity with Weyl-invariant scalar, fermion, Yukawa and gauge sectors. The quadratic structure $(\tilde R-\mu^2|\phi|^2 )^2$ allows the Weyl Goldstone mode to be extracted via a Stueckelberg mechanism independent of the Higgs field. Spontaneous breaking of Weyl gauge symmetry reduces the Weyl quadratic curvature to the Einstein-Hilbert action with a positive cosmological constant, generates a mass term for the Weyl gauge field, and simultaneously produces the Higgs potential $-\mu^2 |\phi|^2+\lambda^2|\phi|^4$---which is otherwise forbidden by the symmetry. Our framefork unifies the Stueckelberg, Higgs and Yukawa mechanisms, reproduces Standard Model mass generation, and predicts additional Higgs-induced contributions to the Weyl gauge field mass, together with a set of Higgs-Weyl couplings. These interactions provide new phenomenological handles, including a vector dark matter candidate, and highlight the geometric origin of mass.}

\begin{document}

\flushbottom
\maketitle

\section{Introduction}
The quest for a grand unified description of fundamental interactions has long been a driving motivation in theoretical physics. In 1918, Weyl proposed a theory aiming to unify electromagnetism and gravity \cite{Weyl:1918ib}. In his framework, the electromagnetic field was interpreted as a non-metric component, endowing spacetime with a non-metric geometric structure. However, this construction faced a fundamental issue: the non-metric term could not yield physically observable quantities, since its interpretation as an electromagnetic field requires an intrinsic gauge symmetry of the action, implying a joint conformal transformation of the Riemannian metric. As a result, this theory failed to define a Weyl-gauge-independent rule for parallel transport of the line element, and was regarded as phenomenologically problematic, a feather that prevented its widespread adoption for several decades. Independently, Kaluza and Klein later proposed a five-dimensional framework unifying electromagnetism and gravity, which laid the foundation for modern extra-dimensional theories. Recent studies combining Weyl geometry with extra dimensions \cite{Arias:2002ew,Liu:2007ku,Liu:2009dt,Guo:2011wr} have further emphasized the role of pure geometric structures in these models. However, in order to resolve the original problem in four dimensions, it is essential to invoke gauge field theories further developed by Yang and Mills, which provide the modern framework to consistently describe interactions with well-defined observables.

The most successful framework of gauge field theory is the Standard Model (SM) \cite{Higgs:1964pj,Salam:1968rm,Glashow:1961tr,Weinberg:1967tq,Englert:1964et,CMS:2012qbp,tHooft:2015vaz}. The Higgs mechanism, proposed in 1964, enabled Yang-Mills theory to describe massive gauge bosons by reducing symmetry, thus providing a concrete link to observed phenomena. The construction of the 
$SU(2)_L\times U(1)_Y$ model in 1968 marked the completion of electroweak theory, which later led to the experimental discovery of the Higgs boson in 2012. The SM unifies three of the four fundamental interactions: electromagnetic, weak, and strong, while a fully verified quantum description of gravity remains elusive. 

Beyond the incompleteness of quantum gravity, the presence of dark matter in the universe further motivates extensions of the SM. Current particle physics
phenomenology research focuses on two central questions: the phenomenology of dark matter and a deeper understanding of electroweak symmetry breaking \cite{Lebedev:2011iq}. Even prior to the Higgs discovery, models had already proposed connections between dark matter and Higgs field \cite{Hall:2009bx,Kanemura:2010sh,Lebedev:2011iq,Hambye:2008bq}. Dark matter candidates, typically referred to as weakly interacting massive particles, have been extensively studied in the post-Higgs era \cite{Chen:2015dea,Tang:2016vch,Krovi:2019hdl,Arcadi:2017kky,Ema:2019yrd,Hu:2023yjn,ATLAS:2022xlo,Arcadi:2019lka,Baldes:2018emh,Berger:2016vxi,Beniwal:2015sdl,Gross:2015cwa,Lopez:2015uma,MAGIC:2016xys,Burikham:2023bil}. A viable dark matter candidate must (1) be absolutely stable or have a lifetime exceeding the age of the universe, and (2) interact very weakly with SM particles. In the framework considered in this work, the Weyl gauge field naturally satisfies these properties and therefore provides a potential vector dark matter candidate.

Guided by these insights, Weyl geometry has recently been developed as a gauge theoretic framework based on Weyl-invariant actions \cite{Zee:1978wi,Fujii:1982ms,Wetterich:1987fm,Cheng:1988zx,Drechsler:1998gy,Nishino:2004kb,Nishino:2009in,deCesare:2016mml,Tang:2020ovf,Oda:2020yyv,Oda:2020cmi,Scholz:2012ev,Scholz:2017pfo,Scholz:2011za,Quiros:2014hua,Wheeler:2018rjb,Scholz:2014kba,BeltranJimenez:2016wxw,Ohanian:2015wva,Jackiw:2014koa}. In Weyl geometry, the non-metricity is defined by
$\tilde \nabla_\mu g_{\alpha\beta}=-q\omega_\mu g_{\alpha\beta}$,
where $\tilde \nabla$ is the Weyl covariant derivative and $\omega_\mu$ is the Weyl gauge field. The Weyl affine connection is
\begin{equation}
\tilde\Gamma^\rho_{\mu\nu}=\Gamma^\rho_{\mu\nu}+\frac{q}{2}[\delta^\rho_\mu\omega_\nu+\delta^\rho_\nu\omega_\mu-g_{\mu\nu}\omega^\rho],
\end{equation}
where $\Gamma^\rho_{\mu\nu}$ is the Levi-Civita connection\footnote{We denote the fields in Weyl geometry with a symbol $`\sim $ ' and the fields in Riemann geometry without this symbol.}. The corresponding curvature tensor is given by
\begin{equation}
    \tilde R^\lambda_{\mu\nu\sigma}=\partial_\nu\tilde\Gamma^\lambda_{\mu\sigma}-\partial_\sigma\tilde\Gamma^\lambda_{\mu\nu}+\tilde\Gamma^\lambda_{\nu\rho}\tilde\Gamma^\rho_{\mu\sigma}-\tilde\Gamma^\lambda_{\sigma\rho}\tilde\Gamma^\rho_{\mu\nu},
\end{equation}
from which we have the following relation:
\begin{equation}
    \tilde R=R-3q\nabla_\mu\omega^\mu-\frac{3}2q^2\omega_\mu\omega^\mu.
\end{equation}
The Weyl gauge transformations take the form
\begin{equation}
    \begin{aligned}
        &g_{\mu\nu}\rightarrow g'_{\mu\nu}=\Omega^2g_{\mu\nu},\\
        &\omega_\mu~\rightarrow~\omega_\mu'=\omega_\mu-\frac2q\partial_\mu \ln\Omega.
    \end{aligned}
\end{equation}
Under these transformations, one can verify that\footnote{The signature we choose is $(+,-,-,-)$.} $\tilde R'=\Omega^{-2}\tilde R$. Then the Weyl-invariant actions can be constructed from these quantities. With the Stueckelberg mechanism \cite{Stueckelberg:1938zz,Ruegg:2003ps}, spontaneous breaking of Weyl gauge symmetry becomes possible \cite{Nishino:2004kb,Nishino:2009in,Tang:2019uex,Tang:2020ovf,Cheng:1988zx,Ghilencea:2018dqd,Ghilencea:2018thl,Ghilencea:2019jux,Ghilencea:2023wwf}, which enables the mass generation for the Weyl gauge field and the producing of an effective action corresponding to our observed universe, thereby perfectly resolving the issue of the extra conformal freedom.

By embedding the SM within this framework, one can consistently realize spontaneous symmetry breaking in a Weyl-invariant setting. Previous attempts \cite{Ghilencea:2018thl,Ghilencea:2021lpa} were restricted to a $\phi^4$ scalar potential due to Weyl gauge symmetry, whereas electroweak symmetry breaking requires the Higgs potential $-\mu^2|\phi|^2+\lambda^2|\phi|^4$. Additionally, in Weyl conformal gravity it was shown that a Higgs potential can emerge from the $\tilde R^2+\beta\tilde R|\phi|^2$ term \cite{Nishino:2009in,Oda:2020cmi}. In these constructions, the Goldstone mode \cite{Ghilencea:2018dqd} associated with Weyl gauge symmetry is identified with the Stueckelberg field, and the subsequent field redefinition involves the Higgs field, which remains as an independent dynamical degree of freedom. In the present framework, the field redefinition is taken to depend only on the Stueckelberg field, so that the Weyl gauge transformation remains independent of the dynamical Higgs degree of freedom. Building on this feature, our construction introduces the non-minimal coupling $(\tilde R-\mu^2|\phi|^2)^2$ within the full $\text{Weyl} \times SU(2)_L \times U(1)_Y$ framework, which simultaneously generates the Einstein-Hilbert action, reproduces the required Higgs potential and induces additional Higgs-Weyl interactions.  Furthermore, upon extending the scalar sector, no direct couplings are induced among different scalar fields within our framework, while such terms can appear in other realizations. Moreover, by embedding the Higgs-Weyl coupling directly into Weyl quadratic gravity, our setup naturally gives rise to a vector dark matter candidate as a byproduct.

After symmetry breaking, this theory reduces the traditional Weyl quadratic gravity to the Einstein-Hilbert action with a positive cosmological constant. Previous studies have applied Weyl quadratic gravity to inflation \cite{Kubo:2018kho,Tang:2020ovf,Harko:2024fnt,Nishino:2004kb,Cai:2021png,Ghilencea:2018thl,Kubo:2018kho,Ferreira:2019zzx,Barnaveli:2018dxo,Lalak:2022wyu}, which show that Weyl-invariant frameworks can capture early universe dynamics. In particular, Ref.~\cite{Tang:2020ovf} also discussed the possibility that the Weyl gauge field can act as a dark matter candidate and explored its connection to the electroweak potential. In the present work, the focus is on the realization of spontaneous symmetry breaking within a Weyl gauge framework and its embedding into the Standard Model sector. The structure admits an extension of the scalar sector, allowing an additional inflationary degree of freedom to be incorporated within the same Weyl-invariant framework without introducing interactions beyond those present in the Higgs sector. This direction is not pursued here, but the framework provides a basis for studying possible connections between early-universe dynamics and electroweak symmetry breaking. Building on these results, our construction can then provide a basis for exploring epochs preceding the emergence of a physically scaled metric, offering a novel perspective on pre-scale cosmology.

In this work we construct a fully $\text{Weyl}\times SU(2)_L\times U(1)_Y$ invariant framework by embedding a Higgs-Weyl non-minimal coupling into quadratic gravity. After symmetry breaking, the theory naturally reduces to the Einstein-Hilbert action with a positive cosmological constant while simultaneously generating the Higgs potential required for electroweak symmetry breaking. In our construction, the Weyl gauge field also emerges as a massive vector field that can serve as a dark matter candidate. This construction thereby provides a geometric origin of mass that unifies the Stueckelberg, Higgs, and Yukawa mechanisms and naturally produces a vector dark matter candidate.

This paper is organized as follows. In Section~\ref{sec2:headings}, we introduce the $\text{Weyl}\times SU(2)_L\times U(1)_Y$ transformation and the corresponding invarient actions, which include the coupling between the Higgs field and the Weyl scalar curvature. In Section~\ref{sec3:headings}, we implement the Stueckelberg, Higgs, and Yukawa mechanisms to sequentially break symmetries. As a result, the Higgs field, and leptons acquire mass, while a cosmological constant and a potential vector dark matter candidate is obtained in the action. In Section~\ref{sec4:headings}, we discuss the possibility of the Weyl gauge field as the dark matter candidate by calculating the phenomenological process relevant to its lifetime and its interactions with the SM. In Section~\ref{sec5:headings}, we give the conclusion and discussion.

\section{$\text{Weyl} \times SU(2)_L\times U(1)_Y$ invariant actions}
\label{sec2:headings}

We begin by specifying the Weyl- and gauge-invariant extension of the electroweak sector that will serve as the starting point of our analysis. 

Weyl gauge symmetry acts simultaneously on the metric and on matter fields. Since our goal is to analyze how the Stueckelberg mechanism connects to the Higgs and Yukawa mechanisms, we restrict attention to the leptonic sector and do not include gluons or quarks\footnote{The quark and gluon sectors follow identically via the same Yukawa structure and decouple from the Weyl gauge field \cite{Ghilencea:2021lpa}.}. Our theory consists of five sectors,
\begin{equation}
\mathcal{L}=\mathcal{L_\text{Geometry}}+\mathcal{L_\text{Gauge Field}}+\mathcal{L_\text{Higgs}}+\mathcal{L_\text{Leptons}}+\mathcal{L_\text{Yukawa}}.
\end{equation}
The crucial ingredient of our construction is the Weyl- and gauge-invariant geometric Lagrangian
\begin{equation}
\mathcal{L_\text{Geometry}}=\sqrt{-g}\text{ }\alpha_1\left(\tilde R-\mu^2|\phi|^2\right)^2,
\end{equation}
where 
$g=\det g_{\mu\nu}$. This term comprises the Weyl geometric sector, a Higgs self-interaction, and a non-minimal coupling between the Weyl scalar curvature and the Higgs field. This unique construction will later serve as the origin of the Higgs potential once the Weyl gauge symmetry is spontaneously broken. To ensure invariance under the $\text{Weyl} \times SU(2)_L\times U(1)_Y$ transformation, we introduce the corresponding gauge fields 
$\omega_\mu$, $\vec A_\mu$ and $B_\mu$.
The Weyl- and gauge-invariant Higgs Lagrangian is
\begin{equation}
\mathcal{L_\text{Higgs}}=\sqrt{-g}\text{ }\left[ \text{}\frac12\left(D_\mu\phi \right)^\dagger D^\mu \phi-\frac{\lambda^4}{4}|\phi|^4\text{}\right],
\label{eq:Higgs}
\end{equation}
with
$D_\mu=\partial_\mu-ig{\vec \tau}\cdot\vec A_\mu/2-i{g'}B_\mu/2-{q}\omega_\mu/2$ and $\phi=\left(\begin{array}{cc}
     \phi^+ \\
     \phi^0
\end{array}\right)$. The Higgs field plays a dual role, entering both the geometric sector and the matter sector.
The invariant gauge-field Lagrangian is 
\begin{equation}
\mathcal{L_\text{Gauge Fields}}=\sqrt{-g}\text{ }\left[-\frac14F_{\mu\nu}F^{\mu\nu}-\frac14A^i_{\mu\nu}A^{i\mu\nu}-\frac14B_{\mu\nu}B^{\mu\nu}\text{}\right],
\end{equation}
where
$F_{\mu\nu}=\partial_\mu\omega_\nu-\partial_\nu\omega_\mu,~B_{\mu\nu}=\partial_\mu B_\nu-\partial_\nu B_\mu,~A^i_{\mu\nu}=\partial_\mu A^i_\nu-\partial_\nu A^i_\mu+g\text{ }\epsilon^{ijk}A^j_\mu A^k_\nu,~i,j,k=1,2,3$.
The lepton sector is given by
\begin{equation}
\mathcal{L_\text{Lepton}}=\sqrt{-g}\text{ }\left[\text{}\psi^\dagger_Li\gamma^ae^\mu_a\nabla^L_\mu\psi_L+\psi^*_Ri\gamma^ae^\mu_a\nabla^R_\mu\psi_R\text{}\right],
\end{equation}
where
$\psi_L=\left(\begin{array}{cc}
     v_L  \\
     l_L
\end{array}\right)$, $\psi_R=l_R$, and the covariant derivative\footnote{According to calculation in Ref. \cite{Ghilencea:2021lpa}, fermions do not couple to the Weyl gauge fields.} is
\begin{equation}
    \begin{aligned}
        \nabla^L_\mu&=\partial_\mu-ig\frac{\vec\tau}{2}\cdot\vec A_\mu+i\frac{g'}{2}B_\mu+\frac12 S^{ab}_\mu \sigma_{ab},\\
\nabla^R_\mu&=\partial_\mu+ig'B_\mu+\frac12 S^{ab}_\mu \sigma_{ab},
    \end{aligned}
\end{equation} 
with the spin connection given by
 $S^{ab}_\mu=e^{\lambda b}\left(\partial_\mu e^a_\lambda-e^a_\nu\Gamma^\nu_{\mu\lambda} \right)$ and $\sigma_{ab}=[\gamma_a,\gamma_b]/4$.
The Weyl-gauge-invariant Yukawa Lagrangian is
\begin{equation}
\mathcal{L_\text{Yukawa}}=\sqrt{-g}~\left[-g_Y\left(\psi_L^\dagger\phi\psi_R+\psi_R^*\phi^\dagger\psi_L \right)\text{}\right],
\end{equation}
where $g_Y$ denotes the Yukawa coupling constant. All components of the action are invariant under $\text{Weyl} \times SU(2)_L\times U(1)_Y$ gauge transformations:
\begin{equation}
\begin{aligned}
&g'_{\mu\nu}=e^{2\theta(x)}g_{\mu\nu},~~{e^\mu_a}'=e^{-\theta(x)}e^\mu_a,\\
&\phi'=\exp\left[-\theta(x)-i\vec\xi(x)\cdot\frac{\vec\tau}{2}-i\frac{\eta(x)}{2}\right]\phi,\\
&\psi_L'=\exp\left[-\frac32\theta(x)-i\vec\xi(x)\cdot\frac{\vec\tau}{2}+i\frac{\eta(x)}{2}\right]\psi_L,\\&\psi_R'=\exp\left[-\frac32\theta(x)+i\eta(x)\right]\psi_R.
\end{aligned}
\end{equation}
It then follows that
$\sqrt{-g'}=e^{4\theta(x)}\sqrt{-g},~\tilde R'=e^{-2\theta(x)}\tilde R$.
The spin connection transforms as
 ${S'^{ab}_\mu}=S^{ab}_\mu+\left(e^a_\mu e^{\nu b}-e^b_\mu e^{\nu a} \right)\partial_\mu\theta(x)$.
The gauge fields transform as
\begin{equation}
\begin{aligned}
&\omega'_\mu=\omega_\mu-\frac{2}{q}\partial_\mu \theta(x),~~B_\mu'=B_\mu-\frac{1}{g'}\partial_\mu\eta(x),\\
&\frac{\vec\tau}{2}\cdot\vec A'_\mu=e^{-i\vec\xi(x)\cdot\frac{\vec\tau}{2}}\left(\frac{\vec\tau}{2}\cdot\vec A_\mu \right)e^{i\vec\xi(x)\cdot\frac{\vec\tau}{2}}+\frac{i}{g}e^{-i\vec\xi(x)\cdot\frac{\vec\tau}{2}}\partial_\mu e^{i\vec\xi(x)\cdot\frac{\vec\tau}{2}},
\end{aligned}
\end{equation}
which confirms that all five Lagrangians are invariant under $\text{Weyl} \times SU(2)_L\times U(1)_Y$ symmetry.
In the next section we discuss how this symmetry is spontaneously broken.

\section{Spontaneous breaking of $\text{Weyl} \times SU(2)_L\times U(1)_Y$ symmetry}
\label{sec3:headings}
\subsection{Breaking the Weyl gauge symmetry by Stueckelberg mechanism}

In order to preserve the full $\text{Weyl} \times SU(2)_L\times U(1)_Y$ symmetry, the Higgs potential in the original Weyl-invariant Lagrangian \eqref{eq:Higgs} is restricted to the quartic form $V(\phi)\propto\phi^4$. However, the SM crucially relies on the Higgs potential: $-\mu^2|\phi|^2+\lambda^2|\phi|^4$, whose explicit mass term enables electroweak symmetry to break. Without a potential of this form, the Higgs mechanism cannot operate, and the subsequent structure of the SM would not emerge. On the other hand, scale invariance forbids the appearance of dimensionful quantities. To recover the proper theory describing our universe, the Weyl gauge symmetry must first be spontaneously broken. This requires invoking the Stueckelberg mechanism, which here plays the role of a gravitational analogue of the Higgs mechanism.

Since the Stueckelberg mechanism is conceptually parallel to the Higgs mechanism, we first extract the Goldstone mode \cite{Ghilencea:2018dqd} from the system. To this end, we examine the Lagrangian 
$\mathcal{L_\text{Geometry}}+\mathcal{L_\text{Higgs}}$. Following the standard strategy used in $f(R)$ gravity, we isolate the higher derivative scalar degree of freedom contained in $(\tilde R-\mu^2|\phi|^2)^2$ by defining
\begin{equation}
\begin{aligned}
F\left(\tilde R,\phi \right)&=\alpha_1\left(\tilde R-\mu^2|\phi|^2\right)^2,\\
G\left(\chi,\tilde R,\phi \right)&=f(\chi)-f'(\chi)\left(\tilde R-\mu^2|\phi|^2+\chi\right),
\end{aligned}
\end{equation}
with 
$f(\chi)=\alpha_1\chi^2$. Then
\begin{equation}
G\left(\chi,\tilde R,\phi \right)=-2\alpha_1\chi\tilde R-\alpha_1\chi^2+\tilde\mu^2\chi|\phi|^2,
\end{equation}
where 
$\tilde \mu^2=2\alpha_1\mu^2$. Thus,
\begin{equation}
\begin{aligned}
\frac{\delta G}{\delta\chi}=0&\Rightarrow\tilde R-\mu^2|\phi|^2=-\chi.
\end{aligned}
\end{equation}
On the on-shell condition 
${\delta G}/{\delta\chi}=0$, one has the equivalence relation:
$\mathcal{L}(G)\Leftrightarrow\mathcal{L}(F)$,  therefore
\begin{equation}
\mathcal{L_\text{Geometry}}+\mathcal{L_\text{Higgs}}=\sqrt{-g}~\left[-2\alpha_1\chi\tilde R-\alpha_1\chi^2+\tilde\mu^2\chi|\phi|^2+\frac12\left(D_\mu\phi \right)^\dagger D^\mu \phi-\frac{\lambda^4}{4}|\phi|^4\right].
\end{equation}
The potential is now
$V(\phi)=-\tilde\mu^2\chi|\phi|^2+{\lambda^2}|\phi|^4/4$,
and the full $\text{Weyl} \times SU(2)_L\times U(1)_Y$ symmetry remains intact, with 
$\chi'=e^{-2\theta(x)}\chi$. We will now use the Stueckelberg mechanism to spontaneously break 
this Lagrangian down to the Einstein-Hilbert term and simultaneously break $V(\phi)=-\tilde\mu^2\chi|\phi|^2+{\lambda^2}|\phi|^4/4$
into the Higgs potential.

Our construction extends the usual Stueckelberg mechanism by incorporating the Higgs field.
We introduce a Weyl scaling factor $\Omega(x)$, parameterized as $\Omega(x)=e^{\theta(x)}$. Here we denote $\chi=\varphi^2$ and fix it according to
$\Omega^2={4\alpha_1}\varphi^2/{M^2}$. Instead of performing a unitary gauge, we perform a conformal transformation that does not transform the $\varphi$ field collectively. The conformal transformation of the metric is 
$\bar g^{\mu\nu}=\Omega^{-2}g^{\mu\nu}$, which yields
\begin{equation}
\begin{aligned}
&\sqrt{-\bar g}=\Omega^4\sqrt{-g},~~\bar e^\mu_a=\Omega^{-1}e^\mu_a,\\
&\bar{S}^{ab}_\mu=S^{ab}_\mu+\left(e^a_\mu e^{\nu b}-e^b_\mu e^{\nu a} \right)\partial_\mu\ln\Omega,\\
&\bar R=\Omega^{-2}R-6~\bar g^{\mu\nu}\nabla_\mu\nabla_\nu\ln\Omega-6~\bar g^{\mu\nu}\nabla_\mu \ln\Omega\nabla_\nu\ln\Omega.
\end{aligned}
\end{equation}
The Higgs field transforms as
\begin{equation}
\bar\phi=\exp[-\theta(x)]\phi,
\end{equation}
and the geometric Lagrangian becomes
\begin{equation}
\begin{aligned}
\mathcal{L_\text{Geometry}}&=\sqrt{-\bar g}\left[-\frac12M^2\bar R+\frac12M^2\left(3q\bar g^{\mu\nu}\nabla_\mu\omega_\nu+\frac32q^2\bar g^{\mu\nu}\omega_\mu\omega_\nu \right)\right.\\&\left.-\frac12M^2\left(6~\bar g^{\mu\nu}\nabla_\mu\nabla_\nu\ln\Omega+6~\bar g^{\mu\nu}\nabla_\mu \ln\Omega\nabla_\nu\ln\Omega \right)\right.\\
&\left.-\frac{M^4}{16\alpha_1}+\frac{\mu^2M^2}{2}|\bar\phi|^2\right].
\end{aligned}
\end{equation}
At this point the $\varphi$ field acquires a kinetic term. By introducing
\begin{equation}
\bar\omega_\mu=\omega_\mu-\frac{2}{q}\nabla_\mu\ln\Omega,
\end{equation}
and using the fact that 
$\Omega$ contains $\varphi$, $\bar\omega_\mu$ now absorbs the Goldstone mode 
$\varphi$. The Weyl gauge field now gets a mass term, thereby breaking the original Weyl gauge symmetry and restoring the missing gauge degree of freedom. The geometric Lagrangian reduces to
\begin{equation}
\mathcal{L_\text{Geometry}}=\sqrt{-\bar g}\text{ }\left[\text{}-\frac{1}{2}M^2\bar R-\frac{M^4}{16\alpha_1}+\frac34M^2q^2\bar\omega_\mu\bar\omega^\mu+\frac{\mu^2M^2}{2}|\bar\phi|^2\text{}\right].
\end{equation} 
The $\varphi$ field has been completely absorbed into 
$\bar\omega_\mu$. We now perform similar field redefinitions for the remaining sectors:
\begin{equation}
\begin{aligned}
&\bar\psi_L=\exp\left[-\frac32\theta(x)\right]\psi_L,\\&\bar\psi_R=\exp\left[-\frac32\theta(x)\right]\psi_R,\\
&\bar B_\mu=B_\mu,~~\frac{\vec\tau}{2}\cdot\vec{\bar{A_\mu}}=\frac{\vec\tau}{2}\cdot\vec A_\mu .
\end{aligned}
\end{equation}
Substituting these redefinitions into the action gives
\begin{equation}
\begin{aligned}
&\mathcal{L_\text{Geometry}}+\mathcal{L_\text{Higgs}}=\sqrt{-\bar g}\text{}\left[\text{}-\frac{1}{2}M^2\bar R-\frac{M^4}{16\alpha_1}+\frac34M^2q^2\bar\omega_\mu\bar\omega^\mu\right.\\
&~~~~~~~~~~~~~~~~~~~~~~~~\left.+\frac12\left(\bar D_\mu\bar \phi \right)^\dagger \bar D^\mu \bar\phi-\left(-\frac{\mu^2M^2}{2}|\bar\phi|^2+\frac{\lambda^2}{4}|\bar\phi|^4 \right)\text{}\right],\\
&\mathcal{L_\text{Gauge Fields}}=\sqrt{-\bar g}\text{ }\left[\text{}-\frac14\bar F_{\mu\nu}\bar F^{\mu\nu}-\frac14\bar A^i_{\mu\nu}\bar A^{i\mu\nu}-\frac14\bar B_{\mu\nu}\bar B^{\mu\nu}\text{}\right],\\
&\mathcal{L_\text{Lepton}}=\sqrt{-\bar g}\text{ }\left[\text{ }\bar \psi^\dagger_Li\gamma^a\bar e^\mu_a\bar \nabla^L_\mu\bar \psi_L+\bar \psi^*_Ri\gamma^a\bar e^\mu_a\bar \nabla^R_\mu\bar \psi_R\text{}\right],\\
&\mathcal{L_\text{Yukawa}}=\sqrt{-\bar g}\text{}\left[\text{}-g_Y\left(\bar \psi_L^\dagger\bar \phi\bar \psi_R+\bar \psi_R^*\bar \phi^\dagger\bar \psi_L \right)\text{}\right].
\end{aligned}
\end{equation}

Thus after transforming to the new field variables, the Weyl gauge symmetry of
$\mathcal{L_\text{Geometry}}+\mathcal{L_\text{Higgs}}$
is spontaneously broken. The Goldstone mode $\varphi$ is absorbed by the Weyl gauge field (and by the redefined fields), thereby eliminating the Weyl gauge freedom. As a result, the original Weyl quadratic scalar curvature term is driven to the Einstein-Hilbert Lagrangian supplemented by a positive scalar-curvature term and a mass term for the Weyl gauge field.

The breaking extends beyond the geometric sector: because the Higgs field couples to the Weyl scalar curvature, the Higgs field spontaneously acquires the Higgs potential,
$V\left(\bar\phi \right)=-{\mu^2M^2}|\bar\phi|^2/2+{\lambda^2}|\bar\phi|^4/4$. In contrast, 
$\mathcal{L_\text{Lepton}}+\mathcal{L_\text{Yukawa}}+\mathcal{L_\text{Gauge Fields}}$
remains Weyl-invariant, as these sectors do not couple to the Weyl scalar curvature. With the Higgs potential now generated, the theory naturally flows into the SM regime. In the next section, we examine how the Higgs and Yukawa mechanisms act in this Weyl gauge framework.
\enlargethispage{2\baselineskip}
\subsection{Higgs mechanism and Yukawa mechanism}

Given that the action is expressed in terms of barred fields, for notational convenience we drop the bars in what follows, i.e. we set 
$`\bar B_\mu\rightarrow B_\mu$' . Noting that
$V(\phi)=-{\mu^2M^2}|\phi|^2/2+{\lambda^2}{4}|\phi|^4/4$,
the potential attains its minimum at 
$|\phi|^2={\mu^2M^2}/\lambda^2$, hence the vacuum expectation value is 
$\langle\phi\rangle\equiv v={\mu M}/{\lambda}$. In the following we parametrize 
$\phi$ around this vacuum expectation value.
At this stage the full Lagrangian
$\mathcal{L_\text{Geometry}}+\mathcal{L_\text{Higgs}}+\mathcal{L_\text{Gauge Fields}}+\mathcal{L_\text{Lepton}}+\mathcal{L_\text{Yukawa}}$
still exhibits an 
$SU_L(2)\times U(1)_Y$ symmetry. We stress that here we do not consider further transformations of the gravitational sector or of the Weyl gauge field. We parametrize the Higgs doublet as
\begin{equation}
\phi=\exp\left[-i\vec\xi(x)\cdot\frac{\vec\tau}{2}-i\frac{\eta(x)}{2}\right]\left(\begin{array}{cc}
     0  \\
     v+h(x) 
\end{array}\right),
\end{equation}
where 
$h(x)$ denotes the physical Higgs excitation. The covariant derivative remains
$D_\mu=\partial_\mu-ig{\vec \tau}\cdot\vec A_\mu/2-i{g'}B_\mu/2-{q}\omega_\mu/2$.
The Lagrangian is still invariant under the local 
$SU_L(2)\times U(1)_Y$ transformation
\begin{equation}
\begin{aligned}
&\phi'=\exp\left[-i\vec\xi(x)\cdot\frac{\vec\tau}{2}-i\frac{\eta(x)}{2}\right]\phi,\\
&\psi_L'=\exp\left[-i\vec\xi(x)\cdot\frac{\vec\tau}{2}+i\frac{\eta(x)}{2}\right]\psi_L,\\
&\psi_R'=\exp\left[i\eta(x)\right]\psi_R,\\
&B_\mu'=B_\mu-\frac{1}{g'}\partial_\mu\eta(x),\\
&\frac{\vec\tau}{2}\cdot\vec A'_\mu=e^{-i\vec\xi(x)\cdot\frac{\vec\tau}{2}}\left(\frac{\vec\tau}{2}\cdot\vec A_\mu \right)e^{i\vec\xi(x)\cdot\frac{\vec\tau}{2}}+\frac{i}{g}e^{-i\vec\xi(x)\cdot\frac{\vec\tau}{2}}\partial_\mu e^{i\vec\xi(x)\cdot\frac{\vec\tau}{2}}.
\end{aligned}
\end{equation}
We now define the two new field coordinates as follows 
\begin{equation}
\begin{aligned}
&\hat B_\mu=B_\mu+\frac{1}{g'}\partial_\mu\eta(x),\\
&\frac{\vec\tau}{2}\cdot\vec{\hat A}'_\mu=e^{i\vec\xi(x)\cdot\frac{\vec\tau}{2}}\left(\frac{\vec\tau}{2}\cdot\vec A_\mu \right)e^{-i\vec\xi(x)\cdot\frac{\vec\tau}{2}}+\frac{i}{g}e^{i\vec\xi(x)\cdot\frac{\vec\tau}{2}}\partial_\mu e^{-i\vec\xi(x)\cdot\frac{\vec\tau}{2}}.
\end{aligned}
\end{equation}
Solving these relations for 
$A_\mu$ and $B_\mu$ and substituting them together with the parametrization of 
$\phi$ back into the Lagrangian yields
\begin{equation}
\begin{aligned}
\mathcal{L_\text{Geometry}}+\mathcal{L_\text{Higgs}}&=\sqrt{-g}\text{}\left[\text{}-\frac{1}{2}M^2R-\left(\frac{M^4}{16\alpha_1}-\frac{\mu^2 M^2v^2}{2}+\frac{\lambda^2 v^4}{4}\right)+\left(\frac34M^2q^2+\frac{1}{8}q^2v^2 \right)\omega_\mu\omega^\mu\right.\\
&\left.+\frac12\partial_\mu h(x)\partial^\mu h(x)+\left(\frac{\mu^2M^2}{2}-\frac{3\lambda^2v^2}{2} \right)h^2(x)+\mu^2 M^2vh(x)\right.\\
&\left.-\lambda^2v^3h(x)-\lambda^2vh^3(x)-\frac{\lambda^2}{4}h^4(x)\right.\\
&\left.+\frac{g^2}{8}(v+h(x))^2\left(\hat A^1_\mu-i\hat A^2_\mu \right)\left(\hat A^{1\mu}+i\hat A^{2\mu} \right)+\frac18(v+h(x))^2\left(g'^2\hat B_\mu-g\hat A^{3}_\mu\right)^2\right.\\
&\left.+\frac{q^2v}{4}h(x)\omega_\mu\omega^\mu+\frac{q^2}{8}h^2(x)\omega_\mu\omega^\mu+\frac{q v}{2}\partial_\mu h(x)\omega^\mu+\frac{q}{2}\partial_\mu h(x)h(x)\omega^\mu
\text{}\right].
\end{aligned}
\end{equation}
Hence the Lagrangian no longer displays the 
$SU_L(2)\times U(1)_Y$ gauge symmetry. It is a procedure equivalent to fixing unitary gauge. Here it is used to make manifest that the Stueckelberg and Higgs treatments are two similar mechanisms, differing only in which Goldstone modes are absorbed. The parametrization is a tool to extract the Goldstone modes. Thus spontaneous symmetry breaking is effectively a field-coordinate transformation: without redefinition the full 
$SU_L(2)\times U(1)_Y$ symmetry is manifest, while after redefinition the symmetry is hidden and the gauge fields acquire masses.
\enlargethispage{2\baselineskip}

Following the standard procedure we introduce the charged and neutral gauge fields,
\begin{equation}
\begin{aligned}
W^-_\mu=\frac{1}{\sqrt2}\left(\hat A^1_\mu+i\hat A^2_\mu \right),~~W^+_\mu=\frac{1}{\sqrt2}\left(\hat A^1_\mu-i\hat A^2_\mu \right),~~Z_\mu=\frac{1}{\sqrt{g^2+g'^2}}\left(g'^2\hat B_\mu-g\hat A^{3}_\mu \right).
\end{aligned}
\end{equation}
In terms of these fields the relevant part of the Lagrangian becomes
\begin{equation}
\begin{aligned}
\mathcal{L_\text{Geometry}}+\mathcal{L_\text{Higgs}}&=\sqrt{-g}\text{}\left[\text{}-\frac{1}{2}M^2R-\left(\frac{M^4}{16\alpha_1}-\frac{\mu^2 M^2v^2}{2}+\frac{\lambda^2 v^4}{4} \right)+\left(\frac34M^2q^2+\frac{1}{8}q^2v^2 \right)\omega_\mu\omega^\mu\right.\\
&\left.+\frac12\partial_\mu h(x)\partial^\mu h(x)+\left(\frac{\mu^2M^2}{2}-\frac{3\lambda^2v^2}{2} \right)h^2(x)+\mu^2 M^2vh(x)\right.\\
&\left.-\lambda^2v^3h(x)-\lambda^2vh^3(x)-\frac{\lambda^2}{4}h^4(x)\right.\\
&\left.+\frac{g^2}{4}(v+h(x))^2W^+_\mu W^-_\mu+\frac{g^2+g'^2}{8}(v+h(x))^2Z_\mu Z^\mu\right.\\
&\left.+\frac{q^2v}{4}h(x)\omega_\mu\omega^\mu+\frac{q^2}{8}h^2(x)\omega_\mu\omega^\mu+\frac{q v}{2}\partial_\mu h(x)\omega^\mu+\frac{q}{2}\partial_\mu h(x)h(x)\omega^\mu
\text{}\right].
\end{aligned}
\end{equation}
Thus the Higgs mechanism breaks the 
$SU_L(2)\times U(1)_Y$ symmetry, endowing 
$W^\pm_\mu,~Z_\mu$ with masses, and additionally giving the Weyl gauge field an extra mass contribution absent in pure Weyl geometry. 

Considering the Yukawa couplings, to generate the fermion masses, the redefinition of fermion fields must take the following form
\begin{equation}
\begin{aligned}
\hat\psi_L(x)&=\exp\left[i\vec\xi(x)\cdot\frac{\vec\tau}{2}-i\frac{\eta(x)}{2}\right]\psi_L(x),\\
\hat\psi_R(x)&=\exp\left[-i\eta(x)\right]\psi_R(x).
\end{aligned}
\end{equation}
Solving for 
$\psi_L(x),~\psi_R(x)$ in terms of the hatted fields and substituting into the Yukawa sector yields
\begin{equation}
\begin{aligned}
\mathcal{L_\text{Yukawa}}=&-\sqrt{-g}\text{ }g_Y\left(\psi_L^\dagger(x) \phi(x) \psi_R(x)+\psi_R^\dagger(x)\phi^\dagger(x)\psi_L(x) \right)\\
=&-\sqrt{-g}\text{ }g_Y\left(\psi_L^\dagger(x) \exp\left[-i\vec\xi(x)\cdot\frac{\vec\tau}{2}-i\frac{\eta(x)}{2}\right]\left(\begin{array}{cc}
     0  \\
     v+h(x) 
\end{array}\right) \psi_R(x)\right.\\
&\left.+\psi_R^*(x)\exp\left[i\vec\xi(x)\cdot\frac{\vec\tau}{2}+i\frac{\eta(x)}{2}\right]\left(\begin{array}{cc}
     0 & v+h(x) 
\end{array}\right)\psi_L(x) \right)\\
=&-\sqrt{-g}\text{ }g_Y\left(\left(\begin{array}{cc}
     \hat v^*_L & \hat l_L^* 
\end{array}\right)\left(\begin{array}{cc}
     0  \\
     v+h(x) 
\end{array}\right) \hat l_R+\hat l_R^*\left(\begin{array}{cc}
     0 & v+h(x) 
\end{array}\right)\left(\begin{array}{cc}
     \hat v_L \\ \hat l_L
\end{array}\right)\right)\\
=&-\sqrt{-g}\text{ }g_Y\left(v\text{ }l^*l+h(x)\text{ }l^*l \right),
\end{aligned}
\end{equation}
where in the last lines we have used standard Dirac notation $l$. Hence, via the Yukawa coupling, the charged-lepton mass is 
$m_l=g_Yv$ with 
$v={\mu M}/{\lambda}$. At this stage, the $SU_L(2)\times U(1)_Y$ symmetry is broken. For the lepton and gauge field sectors, 
$\mathcal{L_\text{Lepton}}+\mathcal{L_\text{Gauge Fields}}$ remain unchanged because they do not couple to the Goldstone modes. 

Collecting all contributions, the total Lagrangian reads
\begin{equation}
\begin{aligned}
\mathcal{L}=&\mathcal{L_\text{Geometry}}+\mathcal{L_\text{Gauge Fields}}+\mathcal{L_\text{Higgs}}+\mathcal{L_\text{Lepton}}+\mathcal{L_\text{Yukawa}}\\
=&\sqrt{-g}\text{ }\left[\text{}-\frac{1}{2}M^2R-\left(\frac{M^4}{16\alpha_1}-\frac{\mu^2 M^2v^2}{2}+\frac{\lambda^2 v^4}{4}\right)+\left(\frac34M^2q^2+\frac{1}{8}q^2v^2\right)\omega_\mu\omega^\mu\right.\\
+&\left.~\frac12\partial_\mu h(x)\partial^\mu h(x)+{\mu^2M^2}h^2(x)+\mu^2 M^2vh(x)
-\lambda^2v^3h(x)-\lambda^2vh^3(x)-\frac{\lambda^2}{4}h^4(x)\right.\\
+&\left.~\frac{g^2}{4}(v+h(x))^2W^+_\mu W^-_\mu+\frac{g^2+g'^2}{8}(v+h(x))^2Z_\mu Z^\mu\right.\\
+&\left.~\frac{q^2v}{4}h(x)\omega_\mu\omega^\mu+\frac{q^2}{8}h^2(x)\omega_\mu\omega^\mu+\frac{q v}{2}\partial_\mu h(x)\omega^\mu+\frac{q}{2}\partial_\mu h(x)h(x)\omega^\mu\right.\\
-&\left.~\frac14\hat F_{\mu\nu}\hat F^{\mu\nu}-\frac14\hat A^i_{\mu\nu}\hat A^{i\mu\nu}-\frac14\hat B_{\mu\nu}\hat B^{\mu\nu} \right.\\
+&\left.~\hat \psi^\dagger_Li\gamma^a e^\mu_a\hat \nabla^L_\mu\hat\psi_L+\hat \psi^*_Ri\gamma^a e^\mu_a\hat\nabla^R_\mu\hat \psi_R-g_Y\left(vl^*l+h(x)l^*l\right)
\right].
\end{aligned}
\end{equation}
Thus the original $\text{Weyl} \times SU(2)_L\times U(1)_Y$ symmetry is broken. Through the Stueckelberg mechanism, the Weyl quadratic gravity sector is spontaneously reduced to the Einstein-Hilbert action with a positive cosmological constant term and the Proca action of the Weyl gauge field. Simultaneously, the Higgs self-interaction potential is converted into the Higgs potential. The resulting Higgs potential then triggers the usual Higgs and Yukawa mechanisms, yielding masses for the gauge bosons and leptons, which results in the breaking of 
$SU(2)_L\times U(1)_Y$ symmetry.

One can further show that the cosmological constant term arises from the spontaneous breaking of the Higgs quartic self-interaction: during the Stueckelberg breaking the Higgs quartic term absorbs the Weyl Goldstone mode 
$\varphi$ and yields the constant term in the effective action. 

Unlike previous constructions, the Weyl gauge field mass in our model has two sources: one arising from the coupling to the Weyl scalar curvature, and another coming from the coupling to the Higgs field. In the final Lagrangian, there are four interaction terms between the Weyl gauge field and the Higgs field. Aside from these terms, the Weyl gauge field has no additional direct couplings to the SM. The masses obtained are
\begin{equation}
    \begin{aligned}
        &m_\omega^2=\frac32M^2q^2+\frac14q^2v^2,~~ m_h^2=2\mu^2M^2,\\
        &m_{W_\pm}^2=\frac12g^2v^2,~~m_Z^2=\frac{v^2}{4}\left(g^2+g'^2 \right),~~m_l=g_Yv.
    \end{aligned}
\end{equation}
Using 
$v={\mu M}/{\lambda}$, we can rewrite these as
\begin{equation}
    \begin{aligned}
    &m_\omega^2=\frac32M^2q^2+\frac{q^2\mu^2M^2}{4\lambda^2},~~m_h^2=2\mu^2 M^2,\\
    &m_{W_\pm}^2=\frac{g^2\mu^2M^2}{2\lambda^2},~~m_Z^2=\frac{\mu^2M^2}{4\lambda^2}\left(g^2+g'^2 \right),~~m_l=\frac{g_Y\mu M}{\lambda}.
\end{aligned}
\end{equation}
We note that the coupling 
$q$ is now tied to the mass of 
$\omega_\mu$, and both these masses and the cosmological constant are related to 
$M=M_P$, which imposes strong constraints on the model parameters. The hierarchy problem (the large separation between 
$m_h=125~$GeV and 
$M_P\approx2.44\times10^{18}~$GeV) is not special to our construction but is generic to models combining the SM with gravity. It may be addressed within extra-dimensional \cite{Arkani-Hamed:1998jmv,Randall:1999ee,Randall:1999vf} or other beyond SM frameworks \cite{Tang:2019uex,Tang:2020ovf,Hu:2023yjn}. In our setting the hierarchy can be parametrized by a small dimensionless 
$\mu$. Our emphasis here is that the present framework links SM masses and the cosmological constant via the Planck scale. The four Weyl-Higgs interaction terms appearing in the final Lagrangian provide phenomenological handles to test the theory. Moreover, under plausible parameter choices the Weyl gauge field can act as a vector dark matter candidate. In this interpretation, dark matter is not merely a geometric artifact but a genuine gauge field emerging from the spontaneous breaking of $\text{Weyl} \times SU(2)_L\times U(1)_Y$ symmetry. In the following section we discuss the viability of the Weyl gauge field as vector dark matter and outline possible observational signatures.

\section{Possibility of Weyl gauge field as dark matter candidate}
\label{sec4:headings}

A particle can be considered a viable dark matter candidate if it satisfies two conditions: (1) it is stable or has a lifetime much longer than the age of the universe, and (2) its interactions with SM particles are sufficiently weak \cite{Tang:2016vch}. In the following, we examine the viability of the Weyl gauge field $\omega_\mu$ as a dark matter candidate. Before doing so, we first determine the relevant parameters in the theory by comparing our theoretical predictions with measured SM quantities. Since we focus on particle-level interactions where gravitational effects are negligible, all subsequent analyses are performed in a Minkowski spacetime background. 

Our model predicts
\begin{equation}
m_h = \sqrt{2}~\mu M, \quad
m_\omega = \sqrt{\frac{3}{2} + \frac{\mu^2}{4 \lambda^2}}Mq,
\end{equation}
with $M = M_P$. Comparison with observational results gives $\mu \approx 3.62 \times 10^{-17}$. Further comparison with the vacuum expectation value of Higgs
\begin{equation}
v = \frac{\mu M}{\lambda} \approx 246~\text{GeV}
\end{equation}
gives $\lambda \approx 0.51$. Since $\mu / \lambda \ll 1$, we further obtain
\begin{equation}
m_\omega \approx \sqrt{\frac{3}{2}} M q,
\end{equation}
so that the coupling $q$ is directly linked to the Weyl gauge boson mass\footnote{Here we choose $q$ to be positive, since the sign of $q$ is irrelevant in this work.}.

In the effective Lagrangian, the interactions between the Weyl gauge field and the Higgs field can be clarified as
\begin{equation}
\begin{aligned}
A_1:\frac{q^2v}{4}h(x)\omega_\mu\omega^\mu, ~~A_2: \frac{q^2}{8}h^2(x)\omega_\mu\omega^\mu, ~~B_1:\frac{q v}{2}\partial_\mu h(x)\omega^\mu, ~~B_2:\frac{q}{2}\partial_\mu h(x)h(x)\omega^\mu.
\end{aligned}
\end{equation}
In standard vector dark matter models, only $A_1$ and $A_2$ are typically considered, since the derivative couplings $B_1$ and $B_2$ can violate an assumed $Z_2$ symmetry and thereby destabilize the dark matter candidate \cite{Hambye:2008bq}. In our model, however, the Weyl gauge field originates from spontaneous symmetry breaking of a geometric gauge symmetry, which constrains the coupling coefficients to be extremely small. As a result, even in the presence of $B_1$ and $B_2$, it is still reasonable to treat $\omega_\mu$ as a viable dark matter candidate.

In this section we only address the origin of the dark matter candidate (the massive Weyl gauge boson) as a consequence of spontaneous symmetry breaking. The computation of the relic abundance, which depends on the cosmological evolution
and on possible non-thermal production mechanisms, is beyond the scope of this work
and will be investigated separately (see \cite{Tang:2019uex,Tang:2020ovf,Hu:2023yjn} for general discussions).
The present analysis only establishes that the particle is stable and extremely weakly coupled,
thereby constituting a viable dark matter candidate. This offers a purely geometric perspective on early universe matter genesis.

We now analyze the viability of $\omega_\mu$ as a dark matter candidate based on the relative magnitudes of
\begin{equation}
m_\omega^2 = \frac{3}{2} M^2 q^2, ~~
m_h^2 = 2 \mu^2 M^2,
\end{equation}
and examine the decay and annihilation processes induced by the four interactions, comparing theoretical predictions with experimental data.

To analyze the lifetime of $\omega_\mu$, we consider three mass hierarchies: $m_\omega > 2 m_h$, $ m_h/2 < m_\omega < 2 m_h$, and $m_\omega < m_h/2$. The $B_1$-induced process $\omega_\mu \leftrightarrow h$ carries a vanishing on-shell tree-level amplitude, so that no physical vertex arises, but this vertex can still influence other processes. To eliminate $B_1$, a field redefinition $\tilde \omega_\mu=\omega_\mu+qv\partial_\mu h/(8m_\omega^2) $ is required. However, the new terms introduced by this procedure are suppressed by $\mathcal{O}\left(v/M^2 \right)$, so we may safely neglect the mixing term directly and the $B_1$ operator will not be considered further.

To analyze the interaction between $\omega_\mu$ and SM particles mediated by $A_1$, we focus on two aspects: whether the Higgs exhibits significant invisible decays and whether $\omega_\mu$ can annihilate via an $s$-channel Higgs. As a representative example, we consider the annihilation process 
\begin{equation}
\omega_\mu \omega_\mu \rightarrow h \rightarrow l \bar l,
\end{equation}
which captures the leading interaction channel between the Weyl gauge field and SM leptons. 
The corresponding annihilation cross section at tree level can be derived as in Ref. \cite{Duch:2017nbe}:
\begin{equation}
    \begin{aligned}
       \langle \sigma v \rangle_{\omega \omega \to \ell \bar{\ell}}=\frac{q^4m_l^2}{96\pi}\frac{\left(1-m_l^2/m_\omega^2 \right)^{3/2}}{\left(4m_\omega^2-m_h^2 \right)^2+m_h^2\Gamma_h^2},
      \end{aligned}
    \end{equation}
where $\Gamma_h$ is the total Higgs width. This process is independent of the Higgs decay width calculations. When $2 m_\omega\approx m_h$, resonance effects enhance the annihilation cross section. For larger $|2 m_\omega-m_h|$, the total Higgs width $3.0^{+2.0}_{-1.5}~$MeV \cite{CMS:2024eka} can be neglected. Under this approximation, the annihilation cross section simplifies to \cite{Lebedev:2011iq}
\begin{equation}
    \begin{aligned}
       \langle \sigma v \rangle_{\omega \omega \to \ell \bar{\ell}}\approx\frac{q^4m_l^2}{96\pi}\frac{\left(1-m_l^2/m_\omega^2 \right)^{3/2}}{\left(4m_\omega^2-m_h^2 \right)^2}.
      \end{aligned}  
    \end{equation}

The $B_2$ term can also mediate annihilation via $hh\rightarrow \omega_\mu$. Here, we focus only on $\omega_\mu$ decay processes because we are primarily concerned with its lifetime. Reverse processes involving $hh\rightarrow \omega_\mu$ are important for relic density and direct detection studies. Our focus here is on whether Weyl gauge bosons generated through spontaneous symmetry breaking can serve as dark matter candidates, while detailed phenomenological studies are left for future work.

The scattering processes induced by $A_2$, such as $\omega_\mu\omega_\mu\rightarrow hh(hh\rightarrow\omega_\mu\omega_\mu)$ and $h\omega_\mu\rightarrow h\omega_\mu$, affect only the dark matter relic abundance but do not contribute to interactions with the SM particles. Therefore, $A_2$ effects are negligible in this work.

Now, we analyze the decay and annihilation process for $\omega_\mu$ according to the three mass hierarchies as follows.

\subsection{The case $m_\omega>2m_h$}

Under this mass hierarchy, we have
\begin{equation}
\sqrt{\frac{3}{2} M^2 q^2} > 2\sqrt{2}~\mu M ,
\end{equation}
which imposes the following lower bound on the Weyl gauge coupling 
$q$ in terms of 
$\mu$:
 \begin{equation}
    \begin{aligned}
    q>\frac{4\mu}{\sqrt{3}}&~.
      \end{aligned}  
    \end{equation}
For $\mu=3.62 \times 10^{-17}$, this leads numerically to 
$q>8.36\times10^{-17}$. In this regime, the decay channel 
$h\rightarrow \omega_\mu\omega_\mu$ is kinematically forbidden, while the decay process 
$\omega_\mu\rightarrow hh$ becomes allowed. This decay is mediated by the interaction term 
$B_2$, yielding the amplitude
\begin{equation}
    \begin{aligned}
       iM_{\omega\rightarrow hh}=i\frac{q}{2}\varepsilon_\mu(k)p^\mu,
      \end{aligned}  
    \end{equation}
where 
$k$ is the momentum of the incoming 
$\omega_\mu$, $\varepsilon_\mu(k)$ is the polarization vector and 
$p_{\mu}$ is the four-momentum of each Higgs boson. 
After summing over its three physical polarizations, the spin-average introduces the usual factor of 
$1/3$ for massive vector boson. The corresponding decay width is
\begin{equation}
    \begin{aligned}
       \Gamma\left({\omega_\mu\rightarrow h~h} \right)=&\frac{|p|}{8\pi m_\omega^2}|\bar{\mathcal M}|^2\\=&\frac{|p|}{24\pi m_\omega^2}|\mathcal M|^2\\
       =&\frac{q^2m_\omega}{768\pi}\left(1-\frac{4m_h^2}{m_\omega^2} \right)^{3/2},
      \end{aligned}  
    \end{equation}
where $\bar{\mathcal M}$ represents the spin-averaged matrix element. Thus, the lifetime of 
$\omega_\mu$ is
\begin{equation}
    \begin{aligned}
       \tau=&\frac{\hbar}{\Gamma\left(\omega_\mu\rightarrow h~h \right)}\\
       =&\frac{768\hbar\pi}{q^2}\frac{m_\omega^2}{\left(m_\omega^2-4m_h^2 \right)^{3/2}}.
      \end{aligned}  
    \end{equation}
Hence the lifetime scales as 
$\tau\propto1/q^2$. In the special case  
$m_\omega=2m_h$, the denominator of the lifetime vanishes. The decay width goes to zero and the lifetime diverges. Since the 
$\omega_\mu$ lifetime decreases with mass, we can determine an upper bound for the mass of a viable dark matter candidate by requiring 
$\tau>t_0\approx6.61\times10^{41}$ GeV$^{-1}(\hbar=c=1)$, the age of the universe. This implies
\begin{equation}
        \frac{1152\hbar\pi M^2}{\left(m_\omega^2-4m_h^2 \right)^{3/2}}>t_0,
   \end{equation}
from which we obtain 
$q<8.37\times10^{-17}$, corresponding to an upper mass bound
$m_\omega<m_{\omega_0}=250.13$ GeV.
Therefore, all 
$\omega_\mu$ with 
$m_\omega>m_{\omega_0}$ are unstable and cannot serve as dark matter candidates. Combining both lower and upper bounds, we find that for the regime 
$m_\omega>2m_h$, the parameter window for a viable dark matter candidate is
$8.36\times10^{-17}<q<8.37\times10^{-17}$,
corresponding to the mass range
$250~$GeV$<m_\omega<250.13$ GeV.

For the annihilation cross section,
substituting the relation 
$m_\omega=\sqrt{3/2}Mq$, we obtain
\begin{equation}
        \begin{aligned}
    \langle \sigma v \rangle_{\omega \omega \to \ell \bar{\ell}}
    =&\frac{m_l^2}{216M^4\pi}\frac{\left(1-m_l^2/m_\omega^2 \right)^{3/2}}{\left(4-m_h^2/m_\omega^2 \right)^2}.
        \end{aligned}
    \end{equation}
Analyzing the monotonic behavior, we find that it increases in the interval 
$m_\omega\in\left(m_l,m_h/2 \right)$, but decreases for 
$m_\omega\in\left(m_h/2,\infty \right)$. Therefore, in the region 
$m_\omega>2m_h$, the annihilation cross section satisfies
\begin{equation}
        \begin{aligned}
         \frac{m_l^2}{216M^4\pi}\frac{\left(1-m_l^2/m_{\omega0}^2 \right)^{3/2}}{\left(4-m_h^2/m_{\omega0}^2 \right)^2} <\langle \sigma v \rangle_{\omega \omega \to \ell \bar{\ell}}<\frac{32m_l^2}{6075M^4\pi}{\left(1-m_l^2/4m_h^2 \right)^{3/2}},
        \end{aligned}
    \end{equation}
which numerically yields
$\langle \sigma v \rangle_{\omega \omega \to \ell \bar{\ell}}\sim~$$\mathcal{O}$($10^{-97}$ cm$^3/$s ).
This value is vastly smaller than the thermal relic annihilation bound 
$10^{-26}$ cm$^3/$s \cite{Lopez:2015uma, MAGIC:2016xys}. Such a large hierarchy is a direct consequence of the highly suppressed
effective couplings between the Weyl gauge field and Higgs field in the present framework.

\subsection{The case $m_h/2<m_\omega< 2m_h$}
\label{4.2:heading}

Within this mass regime, the Weyl gauge coupling satisfies
\begin{equation}
 \frac{\sqrt{2}}2~\mu M<\sqrt{\frac32M^2q^2}<\sqrt{2}~\mu M,
\end{equation}
which implies
$2.10\times10^{-17}<q<8.36\times10^{-17}$.
Given the mass hierarchy in this regime, both decay channels $\omega_\mu\rightarrow hh$ and $h\rightarrow \omega_\mu\omega_\mu$ are kinematically forbidden. Since $m_\omega<2 m_h$, and $\omega_\mu$ couples only to the Higgs sector at tree level, no viable decay channel remains, so $\omega_\mu$ is stable. Consequently, the dominant process is the annihilation channel.

The annihilation cross section falls within
\begin{equation}
        \begin{aligned}
        \frac{32m_l^2}{6075M^4\pi}{\left(1-m_l^2/4m_h^2 \right)^{3/2}}
        <&~ \langle \sigma v \rangle_{\omega \omega \to \ell \bar{\ell}}< \frac{m_l^2}{216M^4\pi}\frac{\left(1-4m_l^2/m_{h}^2 \right)^{3/2}}{m_h^2\Gamma_h^2}.
        \end{aligned}
    \end{equation}
The 
annihilation cross section reaches a maximal value of $\mathcal{O}$ ($10^{-90}$ cm$^3/$s ), which is negligibly small compared to typical weak-scale cross sections. Thus, in the regime $m_h/2<m_\omega< 2m_h$, Weyl gauge field can be fully regarded as a dark matter candidate.

\subsection{The case $m_\omega<m_h/2$}

In this mass regime, we obtain
\begin{equation}
 \sqrt{\frac32M^2q^2}<\frac{\sqrt{2}}2~\mu M,   
\end{equation}
which gives
$q<2.10\times10^{-17}$.
In this regime, the decay channel $\omega_\mu\rightarrow hh$ is kinematically forbidden. As in subsection \ref{4.2:heading}, $\omega_\mu$ is stable. The only relevant tree-level process is then the Higgs decay $h\rightarrow \omega_\mu \omega_\mu$, whose width is given in Ref. \cite{Lebedev:2011iq}:
\begin{equation}
    \begin{aligned}
       \Gamma\left({h\rightarrow\omega_\mu\omega_\mu} \right)=\frac{q^4v^2m_h^3}{256\pi m_\omega^4}\left(1-\frac{4m_\omega^2}{m_h^2}+12\frac{m_\omega^4}{m_h^4} \right)\sqrt{1-4\frac{m_\omega^2}{m_h^2}}.
      \end{aligned}  
    \end{equation}
Substituting the relations of our model yields
\begin{equation}
    \begin{aligned}
       \Gamma\left({h\rightarrow\omega_\mu\omega_\mu}\right)=\frac{\sqrt{2}v^2}{1152\pi \mu}\sqrt{\frac{\mu^2-3q^2}{\mu^2}}\cdot\frac{4\mu^4-12\mu^2q^2+27q^4}{M}.
      \end{aligned}  
    \end{equation}
Within $0\le q\le 2.10\times10^{-17}$, the width decreases monotonically with $q$. For $m_\omega=m_h/2$, it reaches its minimum value $0$, while at $q=0$, it attains the maximum
${\sqrt{2}v^2\mu^3}/({288\pi M})$,
numerically
$\Gamma\left(h\rightarrow\omega_\mu\omega_\mu \right)_{max}=1.84\times10^{-66}$ GeV,
far below the measured total Higgs width
$3.0^{+2.0}_{-1.5}$~MeV \cite{CMS:2024eka}.

For the annihilation process, since this channel kinematically exists only for $m_l<m_\omega< m_h/2$, the cross section satisfies
\begin{equation}
        \begin{aligned}
        0
        <&~ \langle \sigma v \rangle_{\omega \omega \to \ell \bar{\ell}}<\frac{m_l^2}{216M^4\pi}\frac{\left(1-4m_l^2/m_{h}^2 \right)^{3/2}}{m_h^2\Gamma_h^2},
        \end{aligned}
    \end{equation}
also reaching a maximal value of $\mathcal{O}$($10^{-90}$ cm$^3/$s). In this case, the Weyl gauge field is stable at tree level, and its interactions with the SM are vastly weak, making it a viable dark matter candidate as well. 

\section{Conclusion and Discussion}\label{sec5:headings}

In this work, we have developed a theoretical framework that combines the Weyl quadratic gravity model with the electroweak sector of the SM. By introducing a coupling of the Higgs field and the Weyl scalar curvature, a quadratic structure of the Weyl gauge theory is constructed, from which the Goldstone mode is extracted. This allows the Weyl gauge symmetry to be spontaneously broken via the Stueckelberg mechanism, which makes the Weyl gauge field massive. Consequently, the Weyl quadratic gravity is reduced to the Einstein-Hilbert action with a positive cosmological constant and a Proca action for the Weyl gauge field. Simultaneously, the quartic self-interaction of the Higgs field is reduced to the Higgs potential and thus bridges the framework to the SM. Through the Higgs and Yukawa mechanisms, particle masses are generated, while four additional interaction terms between the Weyl gauge field and the broken Higgs field arise. These terms are essential for exploring the Weyl gauge field as a dark matter candidate.

This framework not only enriches the conceptual structure of Weyl geometry but also establishes a unified description linking the Stueckelberg, Higgs and Yukawa mechanisms, providing a geometric perspective on the origin of mass in the early universe. A feature of the present framework is that the Stueckelberg, Higgs and Yukawa mechanisms are realized within a single Weyl-invariant geometric structure. In particular, the Weyl Goldstone mode is absorbed without involving the Higgs degree of freedom, while in existing constructions the Higgs field is often included in the field redefinition. As a result, the Higgs sector retains its standard dynamical role, and the breaking of Weyl gauge symmetry is realized within an independent sector. Beyond its role in mass generation, the Higgs field may also have a significant impact on the cosmological evolution of the early universe, which could be explored in future studies.

When Weyl first proposed Weyl geometry in 1918, it was clear that the presence of Weyl gauge symmetry prevented a correct description of our universe (in the absence of a spontaneous symmetry breaking mechanism at that time), as observables always include a free conformal factor. While historically considered a drawback, within the modern framework of spontaneous symmetry breaking and inflationary cosmology, this conformal freedom can be reinterpreted. The Weyl gauge theory describes the very early universe, where fundamental fields exist while spacetime lacks a physical scale (like a cyclic universe). Its dynamics are governed by the conformal degree of freedom of the Weyl gauge symmetry. Following the Stueckelberg mechanism, the degrees of freedom are reorganized and the metric acquires a physical scale, yielding a symmetry-broken phase governed by the SM gauge symmetry, in which the Higgs and Yukawa mechanisms break the electroweak symmetry and generate masses for the fields. In this perspective, the Weyl gauge theory opens new avenues for modeling the early universe, potentially allowing time and temperature dependent dimensionless functions to describe these symmetry breaking processes in a unified and precise manner.

The discussion in Section~\ref{sec4:headings} regarding the Weyl gauge field as a dark matter candidate complements this picture. Its mass arises predominantly from the Stueckelberg mechanism, and the contribution from the Higgs mechanism is subdominant. In this sense, the Weyl gauge field provides a geometrically motivated realization of vector dark matter, whose properties are directly tied to the structure of symmetry breaking. This offers a natural origin for dark matter. We have also established a rigorous mass range for this candidate, $0$~GeV$<m_\omega<250.13$~GeV. Moreover, the four additional interaction terms arising after symmetry breaking via the Higgs and Yukawa mechanisms may offer novel signatures for direct detection. At the same time, the extremely suppressed annihilation cross section indicates that thermal production alone is insufficient, pointing to additional production mechanisms beyond the thermal paradigm (see \cite{Wang:2022ojc,Ahmed:2020fhc} for related discussions).

\acknowledgments
This work was supported by the National Natural Science Foundation of China (Grants No. 12475056, No. 12247101), the 111 Project under (Grant No. B20063), the Natural Science Foundation of Gansu Province (No. 22JR5RA389, No. 25JRRA799), and Gansu Province's Top Leading Talent Support Plan.

\bibliography{references}

\end{document}